\documentclass[twocolumn]{revtex4-1}

\usepackage{graphicx}
\usepackage{epsfig}

\usepackage[english]{babel}

\usepackage{amsmath,amssymb}
\usepackage{textcomp}

\usepackage{caption}
\usepackage{xcolor}
\usepackage[colorlinks=true]{hyperref}
\hypersetup{linkcolor=[rgb]{0.6 0 0}, citecolor=[rgb]{0 0.4 0}, urlcolor=[rgb]{0 0 0.6}}

\hypersetup{breaklinks=true, hyperindex=true}

\begin{document}

\title{Three-dimensional skyrmion states in thin films of cubic helimagnets}

\author{F. N. Rybakov$^{1,2}$}\thanks{Corresponding author:  f.n.rybakov@gmail.com }
\author{A. B. Borisov$^{1}$}
\author{A. N. Bogdanov$^{2}$} \thanks{  a.bogdanov@ifw-dresden.de }

\address{$^1$Institute of Metal Physics, Ekaterinburg, 620990, Russia}

\address{$^2$IFW Dresden, Postfach 270116, D-01171 Dresden, Germany}
\date{September 20, 2012}

\begin{abstract}
{ A direct three-dimensional minimization of the
standard energy functional shows
that in thin films of cubic helimagnets
chiral skyrmions  are modulated 
along three spatial directions.
The structure of such 3D skyrmions can be thought of
as a superposition of conical modulations along the
skyrmion axis and double-twist rotation in the perpendicular
plane.
Numerical solutions for chiral modulations demonstrate that 
3D skyrmion lattices and helicoids are thermodynamically stable
in a broad range of applied magnetic fields.
Our results disclose a basic physical mechanism underlying
the formation of skyrmion states recently observed in nanolayers of
cubic helimagnets.

}
\end{abstract}

\maketitle

\vspace{5mm}

The Dzyaloshinskii-Moriya (DM) interactions 
arising in noncentrosymmetric magnets as a result 
of their crystallographic handedness,
induce long-range modulations with a fixed sense 
of the magnetization rotation
\cite{Dz64}.
It was also established that
DM interactions provide a unique mechanism 
to stabilize two- and three-dimensional 
(2D and 3D) localized states
(\textit{chiral skyrmions} and \textit{hopfions})
and their bound states (skyrmion lattices)
\cite{JETP89,JMMM94,Rybakov10}.

\begin{figure}[!htb]
\includegraphics[width=\columnwidth]{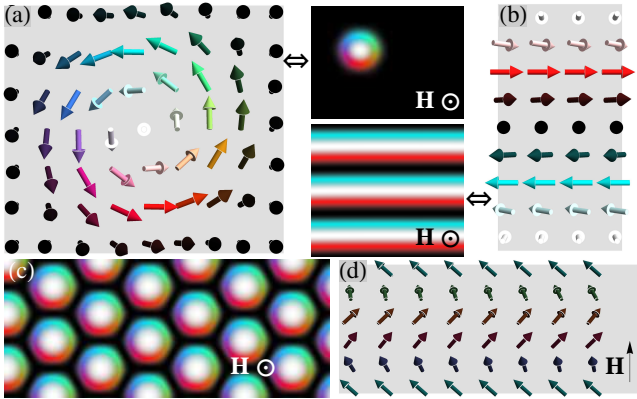}
\caption{ 
(color online). Magnetic configurations in basic
chiral modulations arising in cubic helimagnets:
(a) an isolated skyrmion with double-twist modulations
perpendicular to the applied field, (b) transversally distorted
helices (\textit{helicoids}), (c) a hexagonal lattices
of chiral skyrmions (a), (d) longitudinal distorted helices
(\textit{cones}).
\label{Fig1}
}
\end{figure}

Skyrmion lattices and isolated skyrmions
have been recently discovered in nanolayers 
of cubic helimagnets 
\cite{Yu10,Yu12,Huang12,Wilson} and
in monolayers of magnetic metals with induced
DM interactions \cite{Heinze11}.
Chiral skyrmions are highly mobile 
nanoscale spots of reverse magnetization that 
can be utilized in ultra dense recording technologies
and in the emerging spin electronics
\cite{Kiselev11,Yu12}.
Importantly, in the majority of nonlinear field models 
static multidimensional solitons are
unstable and collapse spontaneously
into topological singularities 
\cite{Derrick64}.
This singles out condensed-matter 
systems  with broken chiral symmetry
as a special class of materials, where chiral
skyrmions can be formed as static, \textit{mesoscopic} 
solitons  and can be manipulated 
in broad ranges of the thermodynamical 
parameters \cite{JETP89,JMMM94}.

The experimental observations of skyrmions and other
chiral modulations \cite{Yu10,Huang12,Karhu12,Wilson}
are in an excellent agreement with results
of the phenomenological theory based on the classical 
Dzyaloshinskii model \cite{Dz64,JMMM94,JPCS11,Karhu12}.
Particularly, theoretical analysis shows
that induced uniaxial anisotropy stabilizes 
skyrmions and helicoids
in thin layers of cubic helimagnets  
\cite{PRB10,Karhu12}. 
In bulk cubic helimagnets,
these chiral modulations exist as metastable
states \cite{JMMM94,PRB10}.
Recent experiments in MnSi and FeGe confirm that
they are stabilize by an induced uniaxial anisotropy
\cite{Karhu12,Huang12,Wilson}.
On the other hand, nothing is known about a role of
induced magnetic anisotropy in mechanically thinned
nanolayers of cubic helimagnets where direct observations
of skyrionic states have been reported
\cite{Yu10,Yu12}.
Generally, the exact magnetic structures of skyrmions
and other modulated states arising in confined cubic
helimagnets are still not resolved and physical mechanisms
underlying their  formation  and stability are  not 
completely understood.

In this paper we provide the first three-dimensional
calculations of chiral modulated states in thin layers
of cubic helimagnet films. We show that below 
a critical thickness, a double-twist skyrmion
gains additional modulations along its axis.
The calculated phase diagram for 3D skyrmions and competing phases 
indicates a broad range of the films thickness, 
where skyrmionic states are thermodynamically stable.

Within the phenomenological theory introduced by Dzyaloshinskii \cite{Dz64}, 
the  magnetic energy density of a cubic non-centrosymmetric ferromagnet 
can be written as \cite{Dz64,Bak80}
\begin{equation}
w =A\,(\mathbf{grad}\,\mathbf{m})^2 + D\,\mathbf{m}\cdot \mathrm{rot}\,\mathbf{m}
-\mathbf{H} \cdot \mathbf{m}\,M.
\label{density}
\end{equation}
$\mathbf{m} = (\sin\theta\cos\psi;\sin\theta\sin\psi;\cos\theta)$ 
is the unity vector along the magnetization $\mathbf{M}$ 
and  $\mathbf{H}$ is the  applied field.
Energy density functional (\ref{density}) includes only the principal 
interactions essential to stabilize skyrmions and helicoids:
the exchange stiffness with constant $A$, DM coupling energy
with constant $D$, and the Zeeman energy. 
We neglect less important energy contributions such as intrinsic and
induced magnetic anisotropies or stray-fields. 
In contrast to common magnetically soft materials where magneto-dipole
interactions play an important role \cite{Hubert98}, 
stray-field effects in noncentrosymmetric magnets  are found to
be weak due to the stabilizing influence of the DM interactions  
\cite{Kiselev11}.

DM interactions favour spatial modulations of $\mathbf{M}$
with a fixed rotation sense \cite{Dz64}. In an infinite sample, 
isolated skyrmions (Fig. \ref{Fig1} a) are described by solutions
of type \cite{JETP89,JMMM94},
\begin{equation}
 \theta = \theta (\rho), \quad \psi = \varphi + \pi/2.
\label{skyrmion}
\end{equation}

Here we write the spatial variables 
in the terms of cylindrical coordinates 
$\textit{\textbf{r}}= (\rho\cos\varphi;\rho\sin\varphi;z)$. 
Eq. (\ref{skyrmion}) describes 2D skyrmions as axisymmetric
tubes with ``double-twist'' modulations
in the $(x,y)$ plane and a homogeneous distribution 
along the skyrmion axis ($z$-axis)
(Fig. \ref{Fig1} a) \cite{Wright89,JMMM94}.
The solutions for one-dimensional modulations include distorted helices
with the propagation direction perpendicular
to the applied field, \textit{helicoids} (Fig. \ref{Fig1} b), and
longitudinal distorted helices modulated along the field, 
\textit{cones} (Fig. \ref{Fig1} d).
The analytical solutions for helicoids describe a gradual unwinding of
chiral coils in an increasing magnetic field \cite{Dz64}.
The equilibrium parameters for the cone phase are given by
\cite{Bak80}
\begin{equation}
 \cos \theta = |\mathbf{H}|/H_D, \quad 
 \psi (z) = \psi_{\mathrm{cone}} (z) \equiv 2\pi z/L_D, 
\label{cone}
\end{equation}
where the period of the helix, $L_D$ and the saturation field, $H_D$ are expressed as
\begin{equation}
  L_D = 4\pi A/|D|,\quad \ H_D = D^2/(2AM).
\label{cone1}
\end{equation}
We introduce the reduced energy density of a modulate phase
as $e = \langle w \rangle /[D^2/(4A)]$ where $\langle w \rangle$ is the energy density
averaged over the modulation period.
Then the  equilibrium energy density of the cone phase can be
written as
\begin{equation}
 e_{\mathrm{cone}} = - \left[1 + (H/H_D)^2 \right]
\label{conee}
\end{equation}

In an infinite sample, the cone (\ref{cone})
corresponds to the global minimum of model (\ref{density})
in the entire magnetic field range below the saturation field 
$H_D$ (\ref{cone1}). Helicoids and skyrmion
lattices only exist as metastable states \cite{PRB10}.

Contrary to lower symmetry noncentrosymmetric
ferromagnets \cite{JETP89},
in cubic helimagnets DM interactions provide
 \textit{three} equivalent 
modulation directions. 
However, 2D chiral skyrmions investigated in 
\cite{JMMM94,JPCS11} have only two (in-plane) modulation
directions while along the  skyrmion
axis the structure remains homogeneous.  
Thus, the question arises: \textit{is it possible to lower 
the skyrmion energy by imposing additional chiral modulations
along the skyrmion axis?}
Our calculations based on rigorous solutions of micromagnetic
equations show that the formation of 3D skyrmions 
 actually occurs in the thickness range below a critical thickness.

As a model  we consider a film of thickness $L$, infinite in
 $x$ and $y$ directions and bounded by parallel 
planes at $z =\pm L/2$.
Numerical solutions for chiral modulations
are derived by direct minimization of
functional (\ref{density}) with free boundary conditions
along the $z$-axis and periodic  boundary conditions 
in the $(x,y)$ plane.
This yields the equilibrium spatially inhomogeneous
distributions of the magnetization vector $\mathbf{m}$
in the layer as functions of the three spatial variables and the two
control parameters, reduced magnetic field, $H/H_D$
and \textit{confinement ratio}, $L/L_D$.

\begin{figure}[!htb]
\includegraphics[width=\columnwidth]{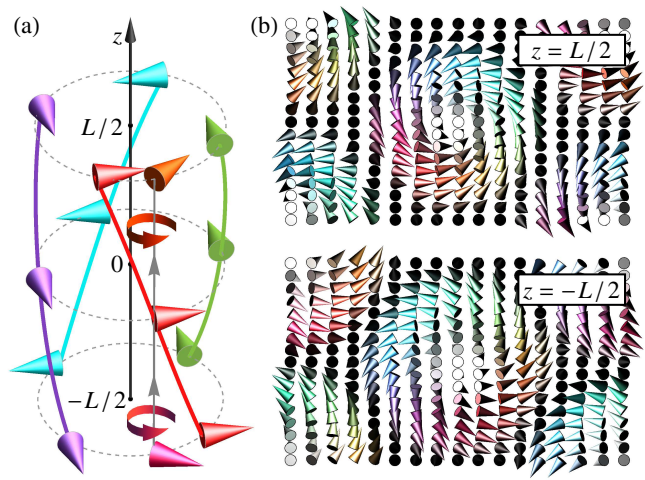}
\caption{
(color online). 3D skyrmion lattice:
(a) Distribution  of the magnetization 
in the skyrmion core demonstrates chiral conical modulations
along the cell axis. For clarity, the sizes along the $z$-axis
are magnified;
(b) Calculated distribution of the magnetization in the
skyrmion cell for top and bottom layers in a film 
with thickness $L/L_D$ = 0.25 and
in the applied field $H/H_D$ = 0.2.
\label{Fig2}
}
\end{figure}

\begin{figure}[!htb]
\includegraphics[width=\columnwidth]{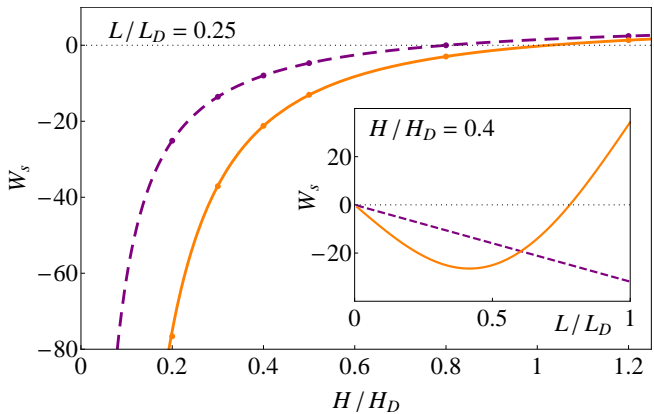}
\caption{ 
(color online). Calculated equilibrium energies of isolated 
skyrmions with ansatz (\ref{3dmode}) (solid line) and for those
that are homogeneous along the $z$-axis (2D skyrmions, dashed line)
as functions of the reduced field $H/H_D$ in 
a film with reduced thickness $L/L_D$ = 0.25. 
Inset shows the equilibrium energies of the skyrmions 
as functions of the reduced thickness at a fixed applied field 
$H/H_D$ = 0.4.
\label{Fig3}
}
\end{figure}

\begin{figure}[!htb]
\includegraphics[width=\columnwidth]{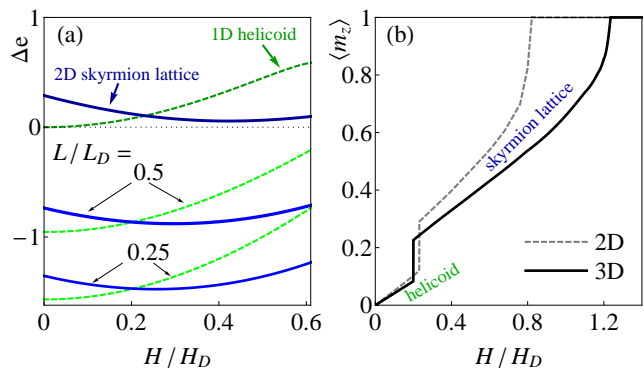}
\caption{
(color online).
(a) The differences $\Delta e$ between the energies of skyrmion lattice
and the cone phase (solid, blue) and between the helicoid and the cone phase
(dashed, green) as functions of the applied field for 2D and 3D modulations.
Functions $\Delta e (H/H_D)$ are plotted for two values of  reduced film
thicknesses $L/L_D$ = 0.25 and 0.5 in comparison with results derived
for 1D helicoids \cite{Dz64} and hexagonal skyrmion lattices homogeneous
along the skyrmion axis (2D skyrmions investigated in \cite{JMMM94}).
(b) Magnetization curves for 3D modulated
structures in a film with reduced thickness $L/L_D$ = 0.25
(solid lines) in comparison with the magnetization curves
for 1D helicoids and 2D skyrmion lattices \cite{Dz64,JMMM94}.
\label{Fig4}
}
\end{figure}

A three-dimensional minimization of nonlinear sigma models needs  huge computing 
efforts and is usually executed on high-performance supercomputers \cite{unwinding}. 
Modern graphics cards offer a promising alternative. 
However, this method requires special hardware-oriented highly parallel algorithms
\cite{PMPP}.
We have developed a special algorithm for CUDA  architecture for NVIDIA graphics cards. 
This architecture has been found to be efficient to solve complex micromagnetic problems 
\cite{MuMax}. 
We use a nonlinear conjugate gradient method 
to minimize energy functional (\ref{density}).
The results of the minimization have been  checked  by 
the compatibility with the  Euler-Lagrange 
equations.
The results of numerical minimization of model 
(\ref{density}) are presented in Figs. 
\ref{Fig2}, \ref{Fig3}, \ref{Fig4} and as video \cite{supplement}.

\textit{Isolated and embedded skyrmions}.
The equilibrium magnetic configurations in 
a 3D skyrmion lattice 
strongly differ from axisymmeteric 2D skyrmions
arising in infinite helimagnets (\ref{skyrmion}).
The magnetization vector 
$\mathbf{m}$ depends on all three spatial components,
$(\rho, \varphi, z)$, and  exhibits complex three-dimensional
modulations ((Fig. \ref{Fig2}) and \cite{supplement}).
It was, however, found  that a simplified ansatz 
\begin{eqnarray}
\theta = \theta(\rho), \quad
\psi (\phi, z)  = \varphi + \pi/2 + \widetilde{\psi} (z)
\label{3dmode}
\end{eqnarray}
with $\widetilde{\psi} (z) = \psi_{\mathrm{cone}}$ (\ref{cone})
provides a nice approximation of the solutions for isolated and
bound 3D skyrmions.

Eq. (\ref{3dmode}) describes double-twist rotations in 
the $(x,y)$ plane and conical modulations
(\ref{cone}) along the skyrmion axis (Fig. \ref{Fig2} a).
Remarkably, the period of conical modulations in the skyrmion
coincides with the helix period, $L_D$. 
We introduce the  reduced equilibrium energy of an isolated skyrmion as
\begin{eqnarray}
W_s = \frac{1}{(A \, L_D)}
\int_0^{2 \pi} d \varphi \int_{-L/2}^{L/2} d z  \int_0^{\infty} 
(w - w_0) \rho d \rho,
\end{eqnarray}
where $w_0 = -H M$ is the energy density of the saturated
state.
In Fig. \ref{Fig3} The energies $W_s$ for  
isolated skyrmions with ansatz (\ref{3dmode}) and
for solutions homogeneous along the skyrmion axis (2D skyrmions)
are plotted as functions of the applied
field in Fig. \ref{Fig3} (main figure) and the reduced film thickness (inset).
These results show that in a broad range of the control parameters
$H/H_D$ and $L/L_D$,  the modulation along the
skyrmion axis reduces its energy.
3D modulated skyrmions should not be confused with three-dimensional
topological chiral solitons (\textit{hopfions}) \cite{Rybakov10}.

\textit{Helicoids and cones}. In thin films, 
helicoids become inhomogeneous along the film thickness
\cite{supplement}. However, we failed to find a simple ansatz
to describe these distortions.
Conical states propagating perpendicular
to the layer surfaces are compatible with the film geometry,
and the free boundary conditions impose no constrictions on these
modulations.  As a result, the solutions for the cone phase
in the film and the equilibrium energy density equal to those
in a bulk sample (\ref{cone}), (\ref{conee}) \cite{Bak80,PRB10}.
Conical states (\ref{cone}) can exist even in films 
with thickness smaller than the helix period ( $L < L_D$)
(\ref{cone1}). In this case  
$\Delta \psi = |\psi (L/2)- \psi (-L/2)| = 2\pi L/L_D < 2\pi$.

The calculated equilibrium energy densities of   helicoids 
($e_{\mathrm{h}})$ and  3D skyrmion lattices 
($e_{\mathrm{sk}})$ have been compared with the cone
energy density ($e_{\mathrm{cone}}$).
The results are plotted as functions 
$\Delta e_{\mathrm{h}} =(e_{\mathrm{h}}-e_{\mathrm{cone}})$ 
and $\Delta e_{\mathrm{sk}} =(e_{\mathrm{sk}}-e_{\mathrm{cone}})$ 
versus the applied field together with the corresponding
results for 1D helicoids and 2D skyrmion lattices arising
in bulk cubic helimagnets \cite{Dz64,JMMM94}
(Fig. \ref{Fig4} a).
In the latter case the cone has the lowest energy in the entire 
region below the saturation field, $H_D$ (\ref{cone}).
This situation  retains in sufficiently thick films 
(see Fig. \ref{Fig3} Inset).
However, the energy balance between the chiral phases  drastically 
changes in thin layers below the critical thickness. 
Here the formation of 3D modulations is resulted in 
a large reduction of the helicoids and skyrmion lattices 
and leads to their energetic advantage over the cone phase.
(Fig. \ref{Fig4} (a)).
It means that in sufficiently thin cubic helimagnets 
films the onset of conical modulations in 
helicoids and skyrmion lattices 
 provides a specific mechanism to
stabilize these chiral modulations
in a broad range of applied magnetic fields
(Fig. \ref{Fig4}). 

An elementary analysis of modulations (\ref{3dmode})
allows to understand basic physical mechanisms underlying 
the formation of 3D skyrmions.  

Inserting ansatz (\ref{3dmode}) into (\ref{density}) yields  
the following expression
\begin{eqnarray}
\tilde{w}&=& A \left( \theta_{\rho}^{2} +\frac{\sin^2 \theta}{\rho^2 }\right)
+ \underbrace{D \cos \widetilde{\psi}  (z) }_{\widetilde{D}(z)}
\left(\theta_{\rho} + \frac{\sin 2 \theta}{2\rho}  \right)
\nonumber \\
&-& H \, M \cos \theta +\underbrace{\sin^2 \theta  \left[ A\widetilde{\psi}_z^2(z) 
- D \widetilde{\psi} _z (z) \right]}_{w_z}  .
\label{isolated1}
\end{eqnarray}

In  functional (\ref{isolated1}) three first terms are responsible for the formation
of  double-twist modulations
in the ($x,y$) plane \cite{JMMM94} and $w_z (z)$ describes modulations along the skyrmion
axis. With conical modes $\widetilde{\psi}(z) = \psi_{\mathrm{cone}}$ (\ref{cone}),
 the function $\widetilde{D}(z)$ and energy density $w_z$ can be written as

\begin{eqnarray}
\widetilde{D} (z) = D \cos \left( 2 \pi z/L_D \right), 
\quad  w_z = -D^2 \sin^2 \theta/(4 A).
\label{cone2}
\end{eqnarray}

Energy density (\ref{isolated1}) includes two different DM terms.
The first, with factor $\widetilde{D} (z)$, favours double-twist rotations 
in the $(x,y)$ plane, the other induces cone-mode modulations along the $z-$ axis.
 The energy gain from the double-twist modulations 
$\propto \widetilde{D}^2 (z)$ reaches the maximum at 
$\widetilde{\psi} (z) = 0$. The cone modulations increase the double-twist energy.
On the other hand, these yield a negative energy contribution $w_z$ (\ref{cone2})
into the total energy. The equilibrium 3D skyrmion patterns are formed as a result 
of the competition between different DM interaction terms.

In the calculated 3D skyrmion textures angle $\widetilde{\psi}$ 
equals zero in the layer center and rotates with a same 
rotation sense to the maximal value at the surfaces, 
$|\widetilde{\psi} (\pm L/2)| =\pi L/L_D$.
Importantly, for $\widetilde{\psi}> \pi/2$ the factor $\widetilde{D} (z)$ 
becomes negative.
It means that cubic helimagnets films
with thickness larger than $L_D/2$ include regions
where function $\widetilde{D} (z)$ becomes negative corresponding
to the double-twist modes with energetically unfavorable 
sense of  rotation. This  explains the existence 
of the critical thickness for 3D modulations and allows
to estimate its value.
As a rule of a thumb, the critical thickness can be estimated as
a half of the helix period $L_D$
(see e.g. Fig. \ref{Fig3} Inset).

The first direct observations of chiral skyrmions 
have been reported in a mechanically thinned  cubic helimagnet (Fe,Co)Si
 with thickness $L= 20$ nm \cite{Yu10}. In this material  $L_D = 90$ nm,
and, thus,  confinement ratio $L/L_D =0.22$. 
This is quite below the critical thickness, and ,thus, 
3D modulations are expected to exist in this film.
In Fig \ref{Fig4} (b) we present the calculated magnetization curve 
for a film with confinement ratio $L/L_D = 0.25$ close to
that of the film investigated by Yu et al. \cite{Yu10}.
This indicates the stability regions for
helicoids and skyrmions, the first order transitions between
helicoid and skyrmion lattices, and the second order transition
of the skyrmion lattice into the saturated state. 

Earlier an alternative mechanism to stabilize chiral skyrmions
in cubic helimagnets has been proposed by Butenko et al. 
\cite{PRB10}.
According to this paper the distortions of a
cone phase imposed by uniaxial magnetic anisotropy
strongly increase  its energy. As a result, chiral skyrmion lattices
and helicoid become thermodynamically stable
in a broad range of applied magnetic fields
(Fig. 3 in \cite{PRB10}).
In our recent studies, the cone modes remain undistorted.
The energetic advantage of 3D skyrmion lattices 
and helicoids is gained because 3D modulations  grant a larger
reduction of the DM energy than single-direction modulations
in cone modes.
New experiments are needed for detailed investigations of
induced magnetic anisotropy and to resolve the exact magnetic
structures in confined cubic helimagnets.
These will allow to understand a role of unaxial distortions
and 3D modulations in the formation of skyrmion states in
these systems.

In conclusion, numerically calculated 3D solutions 
for chiral modulations in nanolayers of cubic heliimagnets 
show that in the films below the critical thickness 
(estimated as a half of the helix period)
chiral skyrmions and helicoids are sufficiently 
inhomogeneous along three spatial directions.
Such 3D modulated skyrmions and helicoids modulated
along the layer thickness
are thermodynamically stable in a broad range 
of applied magnetic fields.

We thank T. Monchesky for critically reading
the manuscript and fruitful advice.

\end{document}